\documentclass[]{fairmeta}

\usepackage{wrapfig}
\usepackage{amsmath}
\usepackage{amssymb}
\usepackage{mathtools}
\usepackage{bm}
\usepackage{algorithm}
\usepackage{algpseudocode}
\usepackage{paralist}
\usepackage[most]{tcolorbox}
\usepackage{booktabs}
\usepackage{array}
\usepackage{tabularx}
\usepackage{ragged2e}
\usepackage{makecell}
\usepackage{multirow}
\usepackage{tikz}
\usepackage{pgfplots}
\usepackage{wrapfig}
\usetikzlibrary{pgfplots.groupplots, matrix}
\pgfplotsset{compat=1.18}
\usepackage{enumitem}
\usepackage{comment}
\usepackage{graphicx}
\usepackage{multirow}
\usepackage{stackengine}
\usepackage{colortbl} 
\usepackage{anyfontsize}

\definecolor{purple}{HTML}{c994c7}
\definecolor{navyblue}{RGB}{30,130,255}
\definecolor{citecolor}{RGB}{30,130,255}
\definecolor{lightgray}{gray}{0.9}
\definecolor{blanchedalmond}{rgb}{1.0, 0.92, 0.8}
\definecolor{cerise}{rgb}{0.871, 0.192, 0.388}

\definecolor{TaskBG}{HTML}{EFE6FF}        
\definecolor{StateBG}{HTML}{F5F5F7}       
\definecolor{ExpertBG}{HTML}{EAF7EA}      
\definecolor{IWMBG}{HTML}{FDECF3}         
\definecolor{SRBG}{HTML}{E6F2FF}          

\newtcolorbox{trainingexample}[2][]{%
  enhanced, breakable, colframe=black!12, colback=white, boxrule=2.5pt,
  arc=2pt, left=0pt, right=0pt, top=0pt, bottom=0pt,
  title={#2}, fonttitle=\bfseries, coltitle=black, #1}

\newcolumntype{L}[1]{>{\RaggedRight\arraybackslash}p{#1}} 
\newcolumntype{Y}{>{\RaggedRight\arraybackslash}X}        

\newtheorem{proposition}{Proposition}
\newtheorem{lemma}{Lemma}

\newcommand{\name}{CoFiRec}

\setlist[itemize]{leftmargin=12pt}

\definecolor{lastauthor}{RGB}{143, 68, 115}

\title{CoFiRec: Coarse-to-Fine Tokenization for Generative Recommendation}

\author[1,\star,\circ]{Tianxin Wei}
\author[1,\star]{Xuying Ning}
\author[2,\dagger]{Xuxing Chen}
\author[1]{Ruizhong Qiu}
\author[3]{Yupeng Hou}
\author[2]{Yan Xie}
\author[2]{Shuang Yang}
\author[2]{Zhigang Hua}
\author[1]{Jingrui He}

\affiliation[1]{University of Illinois Urbana-Champaign}
\affiliation[2]{Meta}
\affiliation[3]{University of California San Diego}

\contribution[\star]{\color{metablue}{\textbf{Equal Contribution}}}
\contribution[\dagger]{\color{lastauthor}\textbf{Core Contributors}}
\contribution[\circ]{Work done at Meta}

\makeatletter
\DeclareFontFamily{T1}{optimistic}{}
\DeclareFontShape{T1}{optimistic}{m}{n}{<->cmr10}{}
\DeclareFontShape{T1}{optimistic}{bx}{n}{<->cmr10}{}
\DeclareFontShape{T1}{optimistic}{m}{it}{<->cmti10}{}
\makeatother

\abstract{
In web environments, user preferences are often refined progressively as users move from browsing broad categories to exploring specific items. However, existing generative recommenders overlook this natural refinement process. Generative recommendation formulates next-item prediction as autoregressive generation over tokenized user histories, where each item is represented as a sequence of discrete tokens. Prior models typically fuse heterogeneous attributes such as ID, category, title, and description into a single embedding before quantization, which flattens the inherent semantic hierarchy of items and fails to capture the gradual evolution of user intent during web interactions. To address this limitation, we propose \textbf{\name}, a novel generative recommendation framework that explicitly incorporates the \textbf{Co}arse-to-\textbf{Fi}ne nature of item semantics into the tokenization process. Instead of compressing all attributes into a single latent space, CoFiRec decomposes item information into multiple semantic levels, ranging from high-level categories to detailed descriptions and collaborative filtering signals. Based on this design, we introduce the \textbf{CoFiRec Tokenizer}, which tokenizes each level independently while preserving structural order. This design naturally reflects the way users refine their preferences, enabling more structured generation. During autoregressive decoding, the language model is instructed to generate item tokens from coarse to fine, progressively modeling user intent from general interests to specific item-level interests. Experiments across multiple public benchmarks and backbones demonstrate that CoFiRec outperforms existing methods, offering a new perspective on structured token design for generative recommendation. Theoretically, we prove that structured tokenization leads to lower dissimilarity between generated and ground truth items, supporting its effectiveness in generative recommendation. Our code is available at \url{https://github.com/YennNing/CoFiRec}.
}

\date{\today}
\correspondence{\email{\{xuyingn2,twei10,jingrui\}@illinois.edu}}

\begin{document}

\maketitle


\section{Introduction}

Generative recommendation (GR) \citep{rajput2023recommender, hua2023index, xu2023openp5, tan2024idgenrec, yue2023llamarec, yang2023generate} has recently emerged as a promising paradigm for sequential recommendation \citep{bert4rec, kang2018self, li2023e4srec}, where user interacted items are tokenized into discrete sequences and modeled autoregressively. This formulation enables GR models to leverage powerful sequence modeling architectures such as Transformers \citep{vaswani2017attention}, while operating over a compact vocabulary of discrete tokens that improves generalization for new items \citep{tan2024idgenrec,rajput2023recommender}, inference efficiency large-scale recommendation, and recommendation quality \citep{rajput2023recommender, lin2024bridging}. These models have demonstrated strong empirical performance across various domains and tasks.

\begin{figure}[t!]
\centering
\includegraphics[width=0.6\linewidth]{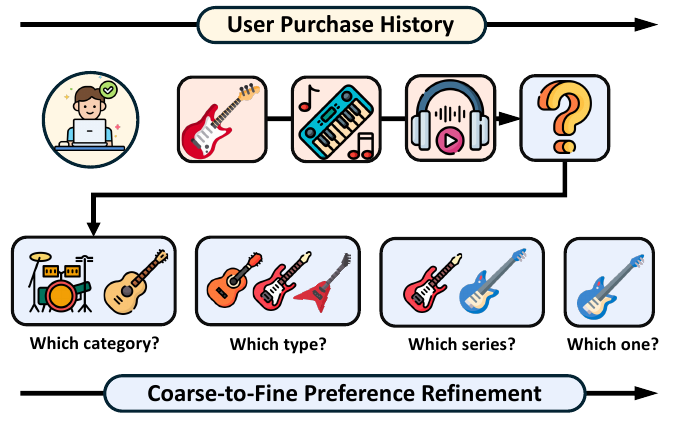}
\vspace{-1em}
\caption{\small Users on the web refine their preferences progressively from category to type to specific series during item exploration.
}
\label{fig:preference_refinement}
\vspace{-1em}
\end{figure}

While GR methods have made substantial progress, they often overlook a key characteristic of real-world recommender systems: the presence of rich, multi-scale semantic structure in item metadata. Existing GR approaches typically flatten all item attributes into a single embedding and encode them using vector quantization, especially residual quantization-based methods like RQ-VAE \citep{lee2022autoregressive}. Although residual quantization introduces an implicit hierarchy in the latent space by encoding residual errors, it fails to capture semantic granularity and interpretability~\citep{rajput2023recommender, tan2024idgenrec, wang2024learnable, yue2023llamarec}. SemID~\citep{hua2023index} provides explicit taxonomy-based semantic indexing, but its design is manually defined and non-learnable, limiting adaptability and discriminative power across datasets.

In contrast, we argue that explicitly modeling the coarse-to-fine nature of item semantics is essential to better align with how users form preferences and navigate item spaces. As illustrated in Figure \ref{fig:preference_refinement}, users typically start from broad categories (e.g., musical instruments), then narrow down to mid-level types (e.g., electric guitars), specific series (e.g., Yamaha Pacifica), and the finally individual item. This multi-level organization is not merely a detail of item annotation, it reflects how users naturally perceive, search for, and refine their preferences over time. Furthermore, collaborative signals \citep{caser, sarwar2001item, he2020lightgcn, zhu2023collaborative, zhang2023collm, zheng2024adapting} such as item co-interaction patterns \citep{sarwar2001item, he2020lightgcn, sarwar2001item} provide an additional layer of fine-grained, behavior-driven semantics that is vital for generative recommendations.

\begin{figure}[t!]
\centering
\includegraphics[width=0.7\linewidth]{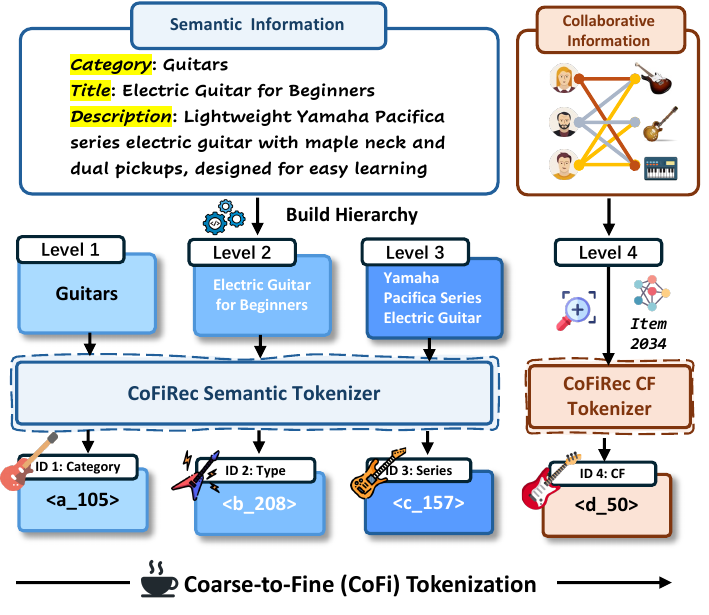}
\vspace{-1em}
\caption{ \small
In \name \ tokenization, item metadata are hierarchically organized (category, title, description) and each level is tokenized independently, with CF signals serve as the most fine-grained tokens.
}
\label{fig:cofirec_token}
\vspace{-1em}
\end{figure}

To address these limitations, we propose \textbf{\name}, a generative recommendation framework that explicitly incorporates the \textbf{Co}arse-to-\textbf{Fi}ne nature of item semantics into both tokenization and generation. As illustrated in Figure~\ref{fig:cofirec_token}, \name~decomposes item metadata and constructs into multiple semantic levels, ranging from high-level categories to fine-grained descriptions, and appends a final item \textit{collaborative filtering (CF) signals} to encode fine-grained behavioral semantics. Unlike prior methods that collapse all item features into a single embedding, \name~retains the hierarchical organization by tokenizing each semantic level independently using coarse-to-fine vector quantization modules in the CoFiRec Tokenizer. This produces a structured token sequence that mirrors the semantic hierarchy of item representation. During generation, the downstream language model is trained to autoregressively decode these tokens in a coarse-to-fine order, reflecting the natural refinement process of user intent. To further enhance generation fidelity and structural alignment, we introduce trainable embeddings that encode both the item identity and the semantic role of each token, enabling the language model to distinguish between levels and maintain consistent semantic conditioning throughout the sequence.

We evaluate our approach on multiple benchmark datasets using LLMs of various scales, including T5~\citep{ni2021sentence}, LLaMA3.2-1B, and LLaMA3.2-3B~\citep{grattafiori2024llama}. Experimental results show that CoFiRec outperforms competitive traditional and generative baselines by an average of 7.27\% across all datasets. Our main contributions are: 

\begin{itemize}
    \item \textbf{Theory-motivated reformulation.} We first theoretically show that hierarchical decoding improves the relevance of recommended items. Motivated by our theory, we propose \name, a novel coarse-to-fine generative recommendation framework to leverage the semantic hierarchy of items in decoding.
    \item \textbf{Coarse-to-fine quantization.} We design the CoFiRec Tokenizer, which uses coarse-to-fine quantization to tokenize multi-granular semantic attributes and collaborative filtering signals into a unified coarse-to-fine token sequence, supporting structured and interpretable generation.
    \item \textbf{Strong performance.} We empirically demonstrate that coarse-to-fine tokenization improves recommendation accuracy by aligning with how users naturally refine their intents.

\end{itemize}

\begin{figure*}[tb!]
\centering
\includegraphics[width=\linewidth]{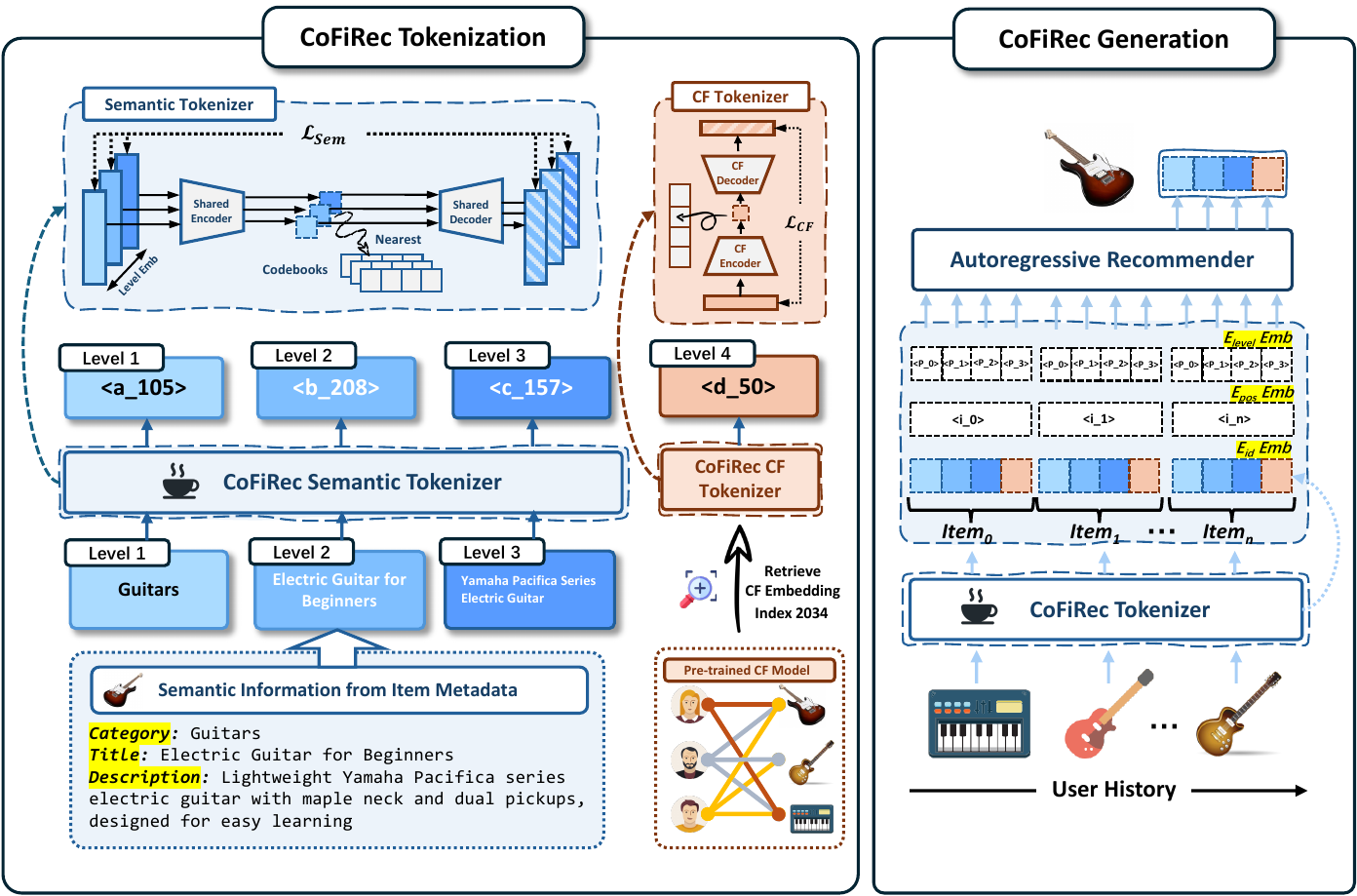}
\caption{ \small
The overall framework of \name, which models multi-level tokenization for coarse-to-fine generative recommendation. 
\textbf{Left:} The CoFiRec Tokenizer decomposes item metadata into hierarchical semantic levels (e.g., category, title, description) and encodes each level using coarse-to-fine VQ modules. Collaborative filtering (CF) signals are tokenized independently from a pre-trained CF embedding via a dedicated CF Tokenizer. 
\textbf{Right:} During generation, an autoregressive language model generates tokens from coarse to fine, conditioned on user history. We inject semantic-level and positional embeddings to help the model identify the role of each token in the structured generation.
}
\label{fig:cofirec}
\end{figure*}

\section{Preliminaries: Generative Recommendation via Discrete Tokens}

\textbf{Problem Setup.} Let $\mathcal{U}$ and $\mathcal{I}$ denote the user and item sets respectively. For each user $u \in \mathcal{U}$, we observe a historical interaction sequence $\mathcal{H}_u = [i_1, i_2, ..., i_T]$, where each $i_t \in \mathcal{I}$. The goal is to predict the next item $i_{T+1}$ that the user is likely to interact with. Generative recommendation models treat this as a sequence modeling task and generate $i_{T+1}$ autoregressively.

\vspace{0.5em}
\noindent\textbf{Tokenization.} To apply language models, each item $i$ must be represented as one token or a sequence of discrete tokens. Prior work uses approaches such as residual quantization \citep{rajput2023recommender} or textual encoding \citep{tan2024idgenrec} to map item embeddings into discrete token sequences.

Let $\bm{e}_i \in \mathbb{R}^d$ denote the dense embedding of item $i$, typically obtained by encoding its flattened multi-level attributes using a pretrained language model. A vector quantization model $\text{VQ}(\bm{e}_i)$ maps $\bm{e}_i$ into a sequence of $K$ discrete tokens $[z_i^{(1)}, z_i^{(2)}, ..., z_i^{(K)}]$, where $z_i^{(k)} \in \mathcal{V}^k$ and $\mathcal{V}^k$ is the codebook at level $k$. The generation process is modeled as:
\begin{equation}
    \mathbb P(z^{(K)},...,z^{(1)}, \mathcal{H}_u) = \prod_{k=1}^{K} \mathbb P(z^{(k)} \mid z^{(<k)}, \mathcal{H}_u),
\end{equation}
where each token $z^{(k)}$ is generated conditioned on the previous token and the user’s interaction history $\mathcal{H}_u$.

\vspace{0.5em}
\noindent\textbf{Limitation.} Existing methods flatten heterogeneous semantic features into the embedding before quantization, discarding the natural hierarchical organization of item attributes. As a result, the generated token sequence lacks interpretability and fails to reflect the coarse-to-fine nature of user preference refinement. Moreover, the standard token-level autoregressive modeling treats each token independently, overlooking the structure across semantic levels.

\section{Method}
\label{sec:method}
In this section, we present the complete methodology of \name, our proposed coarse-to-fine generative recommendation framework. We begin by reviewing the generative recommendation paradigm and review the flaws of existing framework, which frames recommendation as an autoregressive generation task over tokenized item sequences. Next, we introduce multi-level tokenization strategy, detailing the construction of semantic hierarchies and the design of the CoFiRec Tokenizer. Finally, we reformulate the decoding process to align with the coarse-to-fine progression of user intent and describe the training objective for downstream autoregressive modeling. The framework is shown in Figure~\ref{fig:cofirec}.

\subsection{Reformulation \& Theoretical Motivation}
Inspired by the multi-level nature of item semantics and the way users refine their preferences, we reformulate the item recommendation task as a \textit{coarse-to-fine token sequence prediction} problem. Specifically, the model generates a structured sequence of discrete tokens that represent an item, progressing from high-level semantic categories to fine-grained details and behavioral signals. We denote a semantic hierarchy tokenized item as $S_i = [s^{(1)}_i, ..., s^{(K-1)}_i, s^{(K)}_i]$, where the first $K-1$ tokens correspond to progressive semantic levels (e.g., category, title, description), and the final token $s^{(K)}_i$ encodes collaborative filtering (CF) information. We place the CF token last because it captures user-specific interaction signals that are most informative when conditioned on previously decoded semantic content, allowing the model to disambiguate user preferences given coarse-level intent. The overall generation probability is modeled as:
\begin{equation}
    \mathbb P(s^{(K)},...,s^{(1)}, \mathcal{H}_u) = \prod_{k=0}^{K} \mathbb P(s^{(k)} \mid s^{(<k)}, \mathcal{H}_u),
\end{equation}
where each token $s^{(k)}$ is generated in a coarse-to-fine manner, conditioned on all previously generated coarser-level tokens $s^{(<k)}$ and the user’s interaction history $\mathcal{H}_u$, thereby progressively refining the semantic specificity of the predicted item.

Furthermore, we provide a theoretical justification for why hierarchical coarse-to-fine tokenization leads to lower expected dissimilarity compared to independent flat token prediction. 

\vspace{0.5em}\noindent\textbf{Theoretical setup.} We consider two types of generation strategies:

\begin{itemize}
    \item \textbf{Hierarchical decoding setting:} This corresponds to the decoding strategy used in \name, where tokens are generated sequentially from coarse to fine granularity. Let $p$ denote the conditional probability of the correct prediction at each level. We assume that the model is better than random guess (i.e., $p>1/V$, where $V$ is the vocabulary size at each level).
    
    We can view hierarchical decoding as a traversal over a complete $V$-ary tree of depth $K$, where each internal node represents a prefix (partial code) and each edge corresponds to selecting a token from a vocabulary of size $V$. Each item is uniquely represented by a path from the root to a leaf, consisting of $K$ sequential token choices. At each level, the probability of correctly predicting the next token (conditioned on all previous ones) is $p$. We define the dissimilarity $d(i,\hat{\imath})$ between the predicted item $\hat{\imath}$ and the ground-truth item $i$ as half of their distance in the tree.
    
    \item \textbf{Independent decoding setting:} This models the generation process of methods that do not distinguish token granularity. Each token is predicted independently from a flat vocabulary of size $V$, with the same per-token success probability $p$ for fair comparison.
\end{itemize}

\begin{proposition}
\label{prop:proposition}
Let $\mathbb{E}_\textnormal{hier}[d(i,\hat{\imath})]$ and $\mathbb{E}_\textnormal{indep}[d(i,\hat{\imath})]$ denote the expected dissimilarity between the predicted item $\hat{i}$ and the ground-truth item $i$ under hierarchical and independent decoding, respectively. Then, 
\begin{equation}
    \mathbb{E}_\textnormal{hier}[d(i,\hat{\imath})] < \mathbb{E}_\textnormal{indep}[d(i,\hat{\imath})].
\end{equation}
\end{proposition}

\noindent\textit{Proof sketch.} The complete proof 
is deferred to Appendix~\ref{appendix:proof}. 

\vspace{0.5em}
\noindent\textbf{Hierarchical decoding.}
The decoder predicts each token level by level. If $i$ and $\hat\imath$ differ at level $k$, the dissimilarity is $K-k$. The probability of success up to level $k$ is $p^k$, and complete success has probability $p^K$. Hence, by the linearity of expectation, the expected dissimilarity is:
\begin{equation}
\mathbb{E}_\text{hier}[d(i,\hat{\imath})] = K - \sum_{k=1}^{K} p^k = K - p - p^2 - \cdots - p^K.
\end{equation}

\vspace{0.5em}
\noindent\textbf{Independent decoding.}
Each of the $K$ tokens is predicted independently without leveraging the hierarchical structure. This corresponds to randomly sampling a code from the $V^K$ possible leaves, where each leaf (i.e., each item) has equal probability. 
The resulting expected dissimilarity is:
\begin{equation}
\mathbb{E}_\text{indep}[d(i,\hat{\imath})] = \left(K - \sum_{k=1}^{K} \frac{1}{V^k} \right) \cdot \frac{1 - p^K}{1 - \frac{1}{V^K}}.
\end{equation}

Using the fact that the model is better than random guess (i.e., $p > 1/V$), we can show that $\mathbb{E}_\text{hier}[d(i,\hat{\imath})] < \mathbb{E}_\text{indep}[d(i,\hat{\imath})]$. This establishes that hierarchical coarse-to-fine decoding yields a lower expected dissimilarity between predicted and ground-truth items, confirming the theoretical advantage of structured tokenization.

\subsection{Coarse-to-Fine Tokenizer}

\vspace{0.5em}
\noindent\textbf{Hierarchy Construction.} We construct a multi-level semantic hierarchy for each item based on commonly available metadata fields in product recommendation systems, including \textit{category}, \textit{title}, and \textit{description}. These attributes are widely present across most e-commerce and content recommendation platforms and naturally reflect different levels of semantic granularity.  
In our experiments, we use these three levels as the hierarchical structure of item semantics. For a small number of items whose category field is missing, we apply a simple information extraction (IE) procedure to infer category-like terms from their titles or descriptions. 
The constructed $K$-level hierarchy for each item $i$ is thus represented as $[x_i^{(1)}, x_i^{(2)}, ..., x_i^{(K)}]$, where the first $K{-}1$ levels correspond to textual semantics (\textit{category}, \textit{title}, \textit{description}), and the final level $x_i^{(K)}$ encodes \textit{collaborative filtering (CF)} information that captures user–item interaction patterns beyond textual content.

\vspace{0.5em}
\noindent \textbf{CoFiRec Semantic Tokenizer.} Given the constructed semantic hierarchy information, where the first $K{-}1$ levels correspond to progressive semantic levels.
For each semantic level $k \in \{1, 2, \dots, K{-}1\}$, we first obtain an embedding representation $\bm{e}_i^{(k)}$ extracted from a pretrained LLM encoder~\citep{grattafiori2024llama} , which captures contextualized semantic information from the corresponding text field $x_i^{(k)}$. We encode each level independently using a shared semantic encoder $f^{\text{Sem}}$. The encoder is typically implemented as a multi-layer perceptron (MLP) that projects the input feature into the code space. Each level is quantized using a coarse-to-fine vector quantization module $\text{Q}^k$ to capture semantics at different levels:
\begin{equation}
    \bm{h}_i^{(k)} = f^{\text{Sem}}(\bm{e}_i^{(k)}), \quad s_i^{(k)} = \text{Q}^k(\bm{h}_i^{(k)}), \quad \text{for } k = 1, ..., K{-}1.
\end{equation}
All semantic levels share the same encoder and decoder to encourage consistency across levels.

\vspace{0.5em}
\noindent \textbf{CoFiRec CF Tokenizer.} We tokenize CF information as the final $K$-th token. For each item $i$, we retrieve its collaborative embedding $\bm{e}_i^{\text{cf}}$ from a pretrained CF model~\citep{kang2018self}, and encode it with a separate CF-specific encoder and quantize using a dedicated codebook:
\begin{equation}
    \bm{h}_i^{(K)} = f^{\text{CF}}(\bm{e}_i^{\text{cf}}), \quad s_i^{(K)} = \text{Q}^{\text{CF}}(\bm{h}_i^{(K)}) = \text{Q}^{K}(\bm{h}_i^{(K)}).
\end{equation}

\vspace{0.5em}
\noindent\textbf{Vector Quantization Module.} For each semantic level $k$, we maintain a learnable codebook with code embeddings $\{\bm{c}^{(k)}_j\}_{j=1}^{|\mathcal{V}^k|}$. The vector quantization process $\text{Q}^{k}(\cdot)$ assigns the continuous feature $\bm{h}_i^{(k)}$ to its nearest codeword in $\mathcal{V}^k$ via:
\begin{equation}
     s_i^{(k)} \leftarrow \text{Q}^{k}(\bm{h}_i^{(k)}) = \arg\min_{j} \left\| \bm{h}_i^{(k)} - \bm{c}_j^{(k)} \right\|^2,
\end{equation}
where $s_i^{(k)}$ denotes the selected code index for the $k$-th token of item $i$, and $\bm{c}_i^{(k)}$ represents the corresponding selected embedding for item $i$ from the $k$-th codebook. This quantization operation is optimized using the straight-through estimator and a commitment loss, as described above.


\vspace{0.5em}
\noindent\textbf{Optimization.} Each vector quantizer is trained with two objectives: (1) a reconstruction loss to ensure that each decoded representation approximates its original input, and (2) a commitment loss to ensure stable code usage. The total semantic tokenizer loss across all levels is defined as:
\begin{equation}
\mathcal{L}_{\text{Tokenizer}} = \mathcal{L}_{\text{Recon}} + \mathcal{L}_{\text{Commit}},
\end{equation}
where the reconstruction loss is computed between the input embedding $\bm{e}_i^{(k)}$ and its corresponding code embedding $\bm{c}_i^{(k)}$ from the $k$-th codebook at each semantic level $k$:
\begin{equation}
\mathcal{L}_{\text{Recon}} =
\sum_{k=1}^{K-1} 
\underbrace{\left\| \bm{e}_i^{(k)} - g^{\text{Sem}}(\bm{c}_{i}^{(k)}) \right\|^2}_{\mathcal{L}_{\text{Sem}}}
+
\underbrace{\left\| \bm{e}_i^{cf} - g^{\text{CF}}(\bm{c}_i^{(K)}) \right\|^2}_{\mathcal{L}_{\text{CF}}}.
\end{equation}
where $g^{\text{Sem}}$ is a shared semantic decoder that reconstructs continuous semantic embeddings from the corresponding code embeddings, and $g^{\text{CF}}$ is a separate decoder used to reconstruct CF embeddings from CF tokens due to their different representation spaces. The commitment loss is:
\begin{equation}
\mathcal{L}_{\text{Commit}} = \sum_{k=1}^{K} \left\| \text{sg}[\bm{h}_i^{(k)}] - \bm{c}^{(k)}_{i} \right\|^2 + \mu \left\| \bm{h}_i^{(k)} - \text{sg}[\bm{c}^{(k)}_{i}] \right\|^2,
\end{equation}
where $\text{sg}[\cdot]$ is the stop-gradient operator, $\bm{c}^{(k)}_{i}$ is the selected code embedding corresponding to $s_i^{(k)}$ from codebook $\mathcal{V}^k$ at level $k$, and $\mu$ is the balancing coefficient.

After optimizing the CoFiRec tokenizer with $\mathcal{L}_{\text{Tokenizer}}$, each item $i$ is represented by a token sequence $[s_i^{(1)}, ..., s_i^{(K-1)}, s_i^{(K)}]$, which reflects a coarse-to-fine progression from high-level semantics to behaviorally grounded signals.


\subsection{Coarse-to-Fine Autoregressive Generation}

Let $S_u = [S_{i_1}, ..., S_{i_T}]$ denote the tokenized interaction history of user $u$, where the $t$-th item $i_t$ is represented by a coarse-to-fine token sequence $S_{i_t} = [s_{i_t}^{(1)}, s_{i_t}^{(2)}, ..., s_{i_t}^{(K)}]$. We train an autoregressive language model to predict the next item’s token sequence one token at a time, following the same hierarchical order.

\vspace{0.5em}
\noindent\textbf{Embedding Layer.} To help the model distinguish the role of each token, we use three types of learnable embeddings before inputting to the autoregressive language model's encoder:
\begin{equation}
    \bm{e}_{t}^{(k)} = \bm E_\text{id}(s_{i_t}^{(k)}) + \bm E_\text{level}(k) + \bm E_\text{pos}(t),
\end{equation}
where $\bm E_\text{id}$ embeds the token ID, $\bm E_\text{level}(k)$ encodes the semantic level $k$, and $\bm E_\text{pos}(t)$ encodes the item position $t$ in the user history. These embeddings provide the model with structural cues across levels and positions, facilitating more coherent coarse-to-fine generation.

\vspace{0.5em}
\noindent \textbf{Downstream Training Objective.} We adopt a ranking-guided generation loss to enhance token-level discriminability:
\begin{equation}
    \ \mathcal{L}_\text{Rank} = - \sum_{k=1}^{K} \log \frac{\exp(\phi(s_{i_{T+1}}^{(k)})/\tau)}{\sum_{v \in \mathcal{V}^k} \exp(\phi(v)/\tau)},
\end{equation}
where $s_{i_{T+1}}^{(k)}$ is the ground-truth token at level $k$ of the next item at the $T+1$ step to be predicted, $\phi(s_{i_{T+1}}^{(k)})$ is the model’s predicted logit, $\mathcal{V}^k$ is the vocabulary at level $k$, and $\tau$ is a temperature hyperparameter. This objective encourages the model to rank the correct tokens higher than distractors within each semantic level.

\begin{table*}[t!]
\centering
\small
\caption{\small Performance comparison of different methods on the Instruments, Yelp, and Beauty datasets. The best and second-best results are marked in \textbf{Bold} and \underline{Underlined}, respectively. “R@K”: Recall@K; “N@K”: NDCG@K; “BERT4R.”: BERT4Rec; ``LGCN": LightGCN; “IDGenR.”: IDGenRec. “SemID” and “CID” are adapted from~\citep{hua2023index}. The improvements of \name~are statistically significant ($p$~$\ll$~0.05).}
\label{tab:main}
\vspace{-1em}
\resizebox{\textwidth}{!}{
\begin{tabular}{llccccccccccc}
\toprule
\multirow{2}{*}{\textbf{Dataset}} & \multirow{2}{*}{\textbf{Metric}} &
\multicolumn{4}{c}{\textbf{ID-based Recommendation}} &
\multicolumn{7}{c}{\textbf{Generative Recommendation}} \\
\cmidrule(lr){3-6} \cmidrule(lr){7-13}
& & MF & BERT4R. & LGCN & SASRec &
BigRec & SemID & CID & LETTER & TIGER & IDGenR. & \textbf{\name} \\
\midrule
\midrule
\multirow{4}{*}{Instruments}
& R@5  & 0.0479 & 0.0671 & 0.0794 & 0.0751 & 0.0513 & 0.0775 & 0.0809 & 0.0807 & 0.0842 & \underline{0.0843} & \textbf{0.0897} \\ 
& R@10 & 0.0735 & 0.0822 & 0.0100 & 0.0947 & 0.0576 & 0.0964 & 0.0987 & 0.0982 & \underline{0.1035} &
0.0999 & \textbf{0.1118} \\ 
& N@5  & 0.0330 & 0.0560 & 0.0662 & 0.0627 & 0.0470 & 0.0669 & 0.0695 & 0.0704 & 0.0720 & 
\underline{0.0735} & \textbf{0.0752} \\ 
& N@10 & 0.0412 & 0.0608 & 0.0728 & 0.0690 & 0.0491 & 0.0730 & 0.0751 & 0.0760 & 0.0782 & \underline{0.0785} & \textbf{0.0820} \\
\midrule
\multirow{4}{*}{Yelp}
& R@5  & 0.0220 & 0.0186 & 0.0234 & 0.0183 & 0.0154 & 0.0202 & 0.0219 &0.0164 & 0.0203 & \underline{0.0230} & \textbf{0.0252} \\ 
& R@10 & 0.0381 & 0.0291 & \underline{0.0383} & 0.0296 & 0.0169 & 0.0324 & 0.0338 & 0.0259 & 0.0270 & 0.0323 & \textbf{0.0395} \\ 
& N@5  & 0.0138 & 0.0115 & 0.0146 & 0.0116 & 0.0137 & 0.0131 & 0.0140 & 0.0116 & 0.0130 & \textbf{0.0165} & \underline{0.0162} \\
& N@10 & 0.0190 & 0.0159 & 0.0193 & 0.0152 & 0.0142 & 0.0170 & 0.0181 & 0.0147 & 0.0173 & \underline{0.0195} & \textbf{0.0208} \\ 
\midrule
\multirow{4}{*}{Beauty}
& R@5  & 0.0294 & 0.0203 & 0.0305 & 0.0380 & 0.0243 & 0.0393 & \underline{0.0404} & 0.0263 & 0.0349 & 0.0369 & \textbf{0.0444} \\
& R@10 & 0.0474 & 0.0347 & 0.0511 & 0.0588 & 0.0299 & 0.0584 & \underline{0.0597} & 0.0427 & 0.0353 & 0.0550 & \textbf{0.0679} \\
& N@5  & 0.0145 & 0.0124 & 0.0194 & 0.0246 & 0.0181 & 0.0273 & \underline{0.0284} & 0.0161 & 0.0224 & 0.0252 & \textbf{0.0295} \\
& N@10 & 0.0191 & 0.0170 & 0.0260 & 0.0313 & 0.0198 & 0.0335 & \underline{0.0347} & 0.0213 & 0.0283 & 0.0311 & \textbf{0.0370} \\
\bottomrule
\end{tabular}
}
\end{table*}

\section{Experiments}
We conduct comprehensive experiments to evaluate the effectiveness of \name\ in generative recommendation. This section is organized into three parts. First, we describe the \textit{experimental setup}, including datasets, evaluation protocols, and implementation details. Next, we present the \textit{main comparison} with a variety of state-of-the-art baselines to assess overall performance. Finally, we perform detailed \textit{ablation studies} to analyze the contribution of each component, such as hierarchical tokenization, semantic levels, and codebook configurations.

\subsection{Experimental Setup}
\paragraph{Datasets.} 
We evaluate our model on three real-world recommendation datasets from different domains:  
1) \textbf{Instruments} (Musical Instruments) is sourced from the Amazon review datasets~\citep{mcauley2015image} , covering user interactions with music-related products.  
2) \textbf{Beauty} also comes from Amazon reviews, focusing on interactions with beauty products.  
3) \textbf{Yelp}~\citep{yelpdataset} is a popular benchmark dataset containing user reviews of local businesses.  
We follow the data preprocessing procedure in~\citep{kang2018self,li2020time} by removing users and items with fewer than 5 interactions. We also discard items with entirely missing metadata for building semantic hierarchy. To model user behavior sequentially, we adopt the sequential recommendation setting~\citep{quadrana2017personalizing} and use the \textit{leave-one-out} strategy to split the data into training, validation, and test sets~\citep{kang2018self,bert4rec}. For training, we limit each user's interaction history to the most recent 20 items, as done in~\citep{chen2021learning,bert4rec}.


\begin{table*}[t!]
\centering
\begin{minipage}[b]{0.47\textwidth}
    \centering
    \begin{tabular}{lcccc}
        \toprule
        \textbf{Method} & \textbf{R@5} & \textbf{R@10} & \textbf{N@5} & \textbf{N@10} \\
        \midrule
        Reverse Indices & 0.0859 & 0.1047 & 0.0724 & 0.0785 \\
        Random Indices  & 0.0834 & 0.0991 & 0.0725 & 0.0775 \\
        w/o $\bm E_\text{level}$ \& $\bm E_\text{pos}$ & \underline{0.0885} & 0.1072 & \textbf{0.0753} & \underline{0.0813} \\
        w/o $\bm{e}_i^{\text{cf}}$ & 0.0870 & \underline{0.1082} & 0.0740 & 0.0808 \\
        VAR w/ Text     & 0.0675 & 0.0798 & 0.0599 & 0.0639 \\
        \midrule
        \textbf{\name} & \textbf{0.0897} & \textbf{0.1118}  & \underline{0.0752} & \textbf{0.0820} \\
        \bottomrule
    \end{tabular}%
    \subcaption{Ablation study on the Instruments dataset.} 
    \label{tab:ablation}
\end{minipage}
\hfill
\begin{minipage}[b]{0.47\textwidth}
    \centering
    \includegraphics[width=\linewidth]{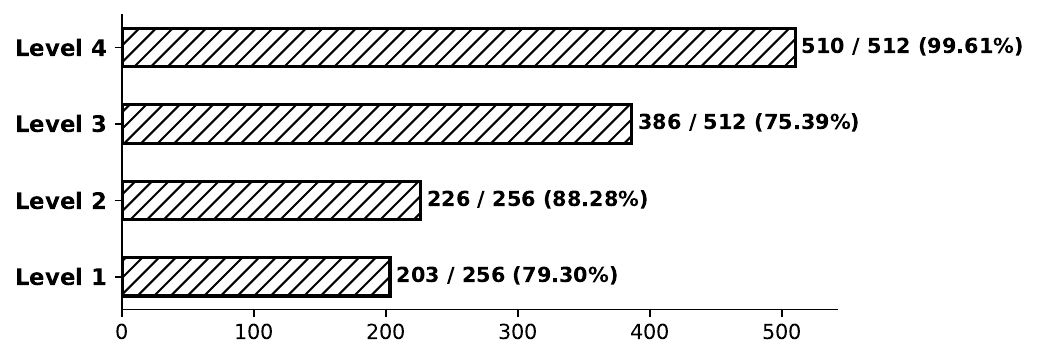}
    \subcaption{Codebook utilization rate of CoFiRec.} 
    \label{fig:enter-label}
\end{minipage}
\caption{Results of the ablation study. (a) shows the ablation study. (b) shows the codebook utilization.}
\vspace{-0.5cm}
\end{table*}

\paragraph{Compared Methods.}
We compare \name\ with a diverse set of strong baselines that fall into two major categories:
For the \textbf{ID-based models}, we include representative approaches such as MF~\citep{rendle2010factorization}, which factorizes the user–item interaction matrix into latent embeddings; BERT4Rec~\citep{bert4rec}, which formulates sequential recommendation as a masked item prediction task using bidirectional transformers; LightGCN~\citep{he2020lightgcn}, which simplifies graph-based collaborative filtering by retaining only the essential message-passing layers; and SASRec~\citep{kang2018self}, which applies self-attention to capture both short- and long-range dependencies in user behavior sequences.
For the \textbf{LLM-based generative models}, we consider BigRec~\citep{bao2025bi}, which directly generates recommendations from item titles; P5-SemID and P5-CID~\citep{hua2023index}, which introduce structured identifiers based on metadata hierarchies and collaborative trees; LETTER~\citep{wang2024learnable}, which combines residual quantization with a diversity constraint to encourage more informative representations; TIGER~\citep{rajput2023recommender}, which represents items as quantized code sequences via RQ-VAE to enable token generation; \citep{wang2024learnable}and IDGenRec~\citep{tan2024idgenrec}, which learns concise textual IDs from human vocabulary.

\paragraph{Evaluation Settings.}
We follow the evaluation protocol of TIGER and report performance using Recall@K and NDCG@K, where $K \in \{5, 10\}$. Both metrics are computed over the ranked list of candidate items for each user. We adopt a leave-one-out evaluation scheme, where the most recent interaction of each user is used for testing, the second most recent for validation, and the rest for training. All models share the same data splits and preprocessing for fair comparison. For each method, we select the checkpoint with the best validation performance for testing. For generative models, beam search with beam size 20 is used during inference following prior work~\citep{rajput2023recommender}.

\paragraph{Implementation Details.}
We represent each item using a token sequence of length \( K = 4 \), where the first three tokens are derived from a semantic tokenizer and the last token comes from a CF tokenizer. For semantic encoding, we extract embeddings from item metadata using a pre-trained \textsc{LLaMA-7B}. For collaborative signals, we adopt a pre-trained SASRec model as the CF encoder, with an embedding dimension of 768. In the first-stage tokenization, we train the discrete tokenizer for 10{,}000 epochs using the AdamW optimizer with a learning rate of 1e-3. Each token position corresponds to a separate codebook, with sizes searched in \{256, 512, 1024\}. Both the encoder and decoder are implemented as multi-layer perceptrons. In the second-stage finetuning, we follow the setup of TIGER~\citep{rajput2023recommender} and use T5 \citep{raffel2020exploring} as the generative backbone. The temperature $\tau$ is set to 1, and $\mu$ is set to 0.25. The learning rate is searched from \{3e-4, 4e-4, 5e-4, 6e-4\}.

\subsection{Overall Performance}
We compare \name\ with both \textit{ID-based} and \textit{generative} recommendation models that adopt different item tokenization strategies, evaluated on three public datasets: Instruments, Yelp, and Beauty. The results are summarized in Table~\ref{tab:main}. 
Overall, generative recommendation models substantially outperform ID-based methods such as MF, BERT4Rec, LightGCN, and SASRec, highlighting the advantage of leveraging textual and semantic signals beyond discrete item IDs. Among generative approaches, IDGenRec performs competitively by encoding item semantics into textual identifiers, demonstrating the benefit of integrating language-based representations. 
Building upon this idea, \name\ further improves performance by introducing a hierarchical coarse-to-fine tokenization scheme that better captures the structured semantics of items. This design preserves multi-level attribute information (e.g., category, title, description, and collaborative signals) and aligns with how users progressively refine their preferences in real-world web scenarios. As shown in Table~\ref{tab:main}, \name\ delivers a 6.4\% improvement over the best-performing baseline, demonstrating its strong generalization ability and effective semantic organization. These results confirm that explicitly modeling the hierarchical structure of item semantics enables more accurate generative recommendation.

\vspace{-0.5em}
\subsection{Ablation Study}
We conduct ablation experiments in Table~\ref{tab:ablation} to examine the contribution of each proposed component in CoFiRec. 
\begin{itemize}[leftmargin=*,itemsep=2pt]
    \item \textit{Effect of token structure:} to assess the importance of the coarse-to-fine token structure, we test two variants: reversed token order and random permutation. Both lead to significant performance drops, with the random order performing worst. This indicates that preserving semantic structure in tokenization facilitates more accurate generation in autoregressive models. 
    
     \item \textit{Effect of collaborative filtering integration.} We replace the final CF token $\bm{e}_i^{\text{cf}}$ with an additional semantic-level token quantized from title and description. This modification degrades performance, showing that incorporating collaborative signals at the finest level provides essential disambiguation for semantically similar items. The CF token complements textual semantics by capturing user–item co-occurrence patterns.
    
    \item \textit{Effect of embedding components in the generation phase:} removing the level embedding $\bm{E}_\text{level}$ and position embedding $\bm{E}_\text{pos}$ causes a clear decline in performance, demonstrating that explicitly encoding token type and position helps the model distinguish semantic roles more effectively.
    
    \item \textit{Comparison with alternative hierarchy construction method:} we also adapt VAR~\citep{tian2024visual} by applying multi-scale pooling over textual embeddings to construct a hierarchy from the embedding space. This yields inferior results compared to our explicit semantic decomposition, likely due to semantic loss introduced by pooling in the one-dimensional embedding space.
\end{itemize}

\begin{figure*}[t!]
  \centering
  \begin{minipage}[t]{0.47\textwidth}
    \centering
    \includegraphics[width=\textwidth]{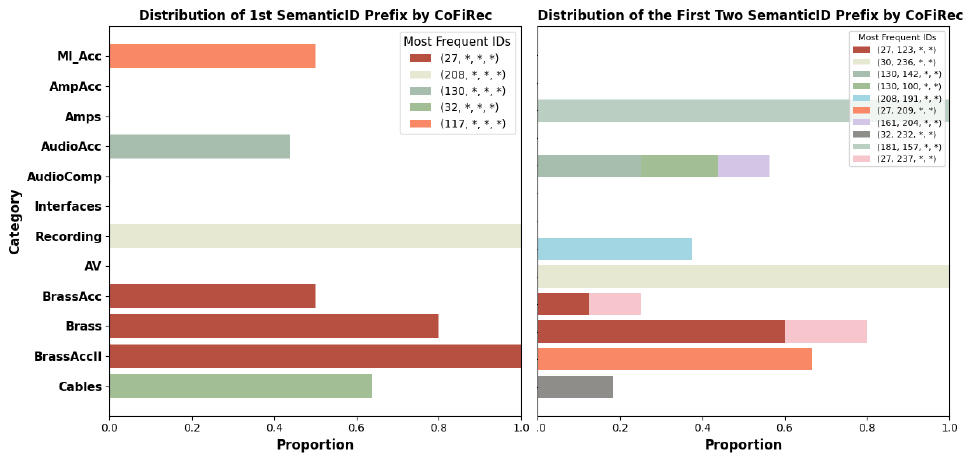}
    \vspace{-20pt}
    \caption*{\small (a) \name}
  \end{minipage}
  \begin{minipage}[t]{0.47\textwidth}
    \centering
    \includegraphics[width=\textwidth]{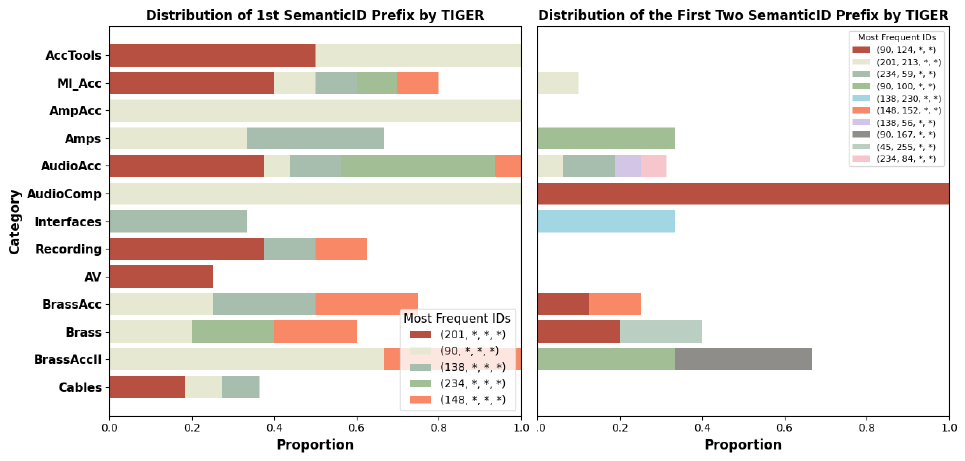}
    \vspace{-20pt}
    \caption*{\small (b) TIGER}
  \end{minipage}
  \vspace{-1em}
  \caption{Distributions of semantic IDs on the Instruments dataset. 
The left panel shows the distribution of \name, where both the first and first-two ID prefixes align closely with category semantics and form clear hierarchical structures, 
while the right panel shows the distribution of TIGER, where prefixes are scattered across categories with weaker semantic consistency.}
  \label{fig:semantic-dist-compare}
\end{figure*}

\section{Analysis}
To gain deeper insights into \name's behavior, we conduct analyses from several perspectives. We study the \textit{impact of codebook sizes}, examine the \textit{semantic structure of ID distributions}, and present a \textit{case study on cold-start generalization}. We further explore \textit{automatic hierarchy construction} from minimal metadata and evaluate \textit{backbone adaptability} across different LLM architectures. 

\subsection{Impact of Codebook Sizes}
CoFiRec uses separate codebooks at each semantic level. Code utilization as shown in Tabel~\ref{fig:enter-label} gradually increases from coarse to fine levels, indicating that deeper layers encode more detailed item semantics and therefore require larger code capacity for discrimination. This observation is consistent with the intuition that higher-level categories represent broader semantic clusters, while deeper levels encode finer-grained and more distinctive item semantics.
Moreover, \name\ naturally reduces code collisions through its hierarchical semantic decomposition and the integration of collaborative signals, as shown in Appendix~\ref{app:code_collison}. This design contrasts with the RQ-VAE tokenizer in TIGER, which relies on appending additional tokens to resolve code overlaps, often leading to redundancy and semantic drift.  
We further analyze the impact of different codebook configurations, as shown in Table~\ref{tab:codebook}. The medium configuration (256, 256, 512, 512) achieves the best overall performance, suggesting that fine-grained levels benefit from moderately larger codebooks to distinguish similar items. However, excessively large codebooks, e.g., (256, 256, 512, 1024), may over-partition the latent space, fragmenting semantics and slightly degrading performance. These results underscore the importance of balanced capacity allocation across semantic levels for achieving both expressiveness and stability in generative recommendation.

\begin{table}[t!]
\centering
\small
\caption{ Impact of different codebook sizes at each level on the Instruments dataset.}
\label{tab:codebook}
\vspace{-1em}
\resizebox{0.7\linewidth}{!}{%
\begin{tabular}{lccccc}
\toprule
\textbf{Size} & \textbf{Level Size} & \textbf{R@5} & \textbf{R@10} & \textbf{N@5} & \textbf{N@10} \\
\midrule
Small  & [256,256,256,256]  & 0.0860 & 0.1078 & 0.0725 & 0.0795 \\
Medium & [256,256,512,512]  & \textbf{0.0893} & \textbf{0.1111} & \textbf{0.0751} & \textbf{0.0822} \\
Large  & [256,256,512,1024] & 0.0867 & 0.1065 & 0.0740 & 0.0804 \\
\bottomrule
\end{tabular}
}
\end{table}

\subsection{Qualitative Analysis of ID Distribution of CoFiRec Tokenization}
To better understand how well different tokenization methods preserve semantic structure, we visualize the distribution of semantic IDs across item categories for both \name\ and TIGER on the Instruments dataset. Figure~\ref{fig:semantic-dist-compare} shows the distribution of the most frequent first-level semantic ID prefixes. We observe that under \name, items sharing the same prefix are highly aligned with specific high-level categories, for example, items in the same coarse-grained category tend to share the same first token. In contrast, TIGER's IDs are more scattered, with no clear pattern of prefix-category alignment, indicating weaker semantic coherence in its token assignments.
Figure~\ref{fig:semantic-dist-compare} further examines the two-level prefix distributions. It reveals that even when items share the same first-level token (coarse semantics), they are further differentiated at the second level, capturing fine-grained category distinctions. This hierarchical organization is especially evident in \name, but much less structured in TIGER.
These results confirm that \name’s tokenizer preserves semantic hierarchy through a coarse-to-fine structure, grouping similar items under shared prefixes while capturing finer distinctions in deeper tokens, thereby enhancing both semantic coherence and interpretability.

\subsection{Case Study on Cold-Start Generalization}
To evaluate the generalization ability of our model in cold-start scenarios, we construct a test set containing items that do not appear in the training data. Cold items are defined as those in the bottom 2\% of item frequency, measured by their occurrences as the last item in user interaction sequences. The tokenizer is trained solely on the training items and then used to infer semantic IDs for these unseen cold-start items.
As shown in Table~\ref{tab:cold-start}, \name\ consistently outperforms TIGER on both warm and cold subsets, achieving a 4.7\% improvement in the warm-start case and a larger 10.0\% gain under the cold-start setting. This clearly demonstrates that \name\ not only maintains strong performance when sufficient user–item interactions are available, but also generalizes more effectively to unseen or infrequent items. We attribute this advantage to its coarse-to-fine tokenization design, which captures structured semantic dependencies among item attributes. By leveraging this hierarchical organization, \name\ effectively transfers semantic patterns from known items to unseen ones, achieving robust performance even under challenging cold-start conditions.

\subsection{\name\ with Automatically Constructed Hierarchies}

Although item hierarchical metadata is widely available in most recommendation datasets, real-world scenarios may lack such structured information. To explore whether meaningful hierarchies can still be automatically derived under this condition, we investigate the ability to generate hierarchical semantics from minimal item metadata. Due to budget constraints, we randomly select 1\% users and their corresponding 2,258 interacted items from the Instruments dataset. Using GPT-4o, we construct 4-level semantic hierarchies based solely on each item's title and description, without access to any structured metadata such as category or brand (examples are provided in Appendix~\ref{app:prompt}).
We denote this variant as AHC (Automatic Hierarchy Construction). While TIGER and \name\ rely on complete metadata fields, AHC leverages only the GPT-generated hierarchies to form its tokenization structure. As shown in Appendix~\ref{app:AHC}, AHC achieves competitive performance, approaching the results of full-metadata \name. This demonstrates that large language models can effectively infer coarse-to-fine semantic relationships from unstructured text, serving as a viable alternative when explicit metadata is unavailable. These findings highlight the broader potential of integrating LLM-based reasoning with generative recommendation for automatic semantic enrichment.

\begin{table}[t!]
\centering
\small
\vspace{-2mm}
\caption{ Performance comparison under warm and cold-start scenarios on the Instruments dataset.}
\label{tab:cold-start}
\vspace{-1em}
\resizebox{0.62\linewidth}{!}{
\begin{tabular}{llcccc}
\toprule
\textbf{Scenario} & \textbf{Method} & \textbf{R@5} & \textbf{R@10} & \textbf{N@5} & \textbf{N@10} \\
\midrule
\multirow{2}{*}{Warm} 
& TIGER & 0.0845 & 0.1038 & 0.0723 & 0.0784 \\
& \name~{\scriptsize\textcolor{red!60!black}{(+4.7\%)}} 
& \textbf{0.0889} & \textbf{0.1101} & \textbf{0.0747} & \textbf{0.0816} \\
\midrule
\multirow{2}{*}{Cold} 
& TIGER & 0.0654 & 0.0801 & 0.0673 & 0.0688 \\
& \name~{\scriptsize\textcolor{green!50!black}{(+10.0\%)}} 
& \textbf{0.0707} & \textbf{0.0943} & \textbf{0.0690} & \textbf{0.0767} \\
\bottomrule
\end{tabular}
}
\end{table}

\begin{table}[t!]
\vspace{-1em}
\centering
\caption{Performance comparison of CoFiRec and TIGER tokenization under different backbone models on the sampled Instruments dataset.}
\label{tab:backbone}
\vspace{-1em}
\resizebox{0.58\columnwidth}{!}{
\begin{tabular}{llcccc}
\toprule
\textbf{Backbone} & \textbf{Method} & \textbf{R@5} & \textbf{R@10} & \textbf{N@5} & \textbf{N@10} \\
\midrule
\multirow{2}{*}{T5} 
& TIGER & 0.0444 & 0.0484 & \textbf{0.0293} & 0.0307 \\
& \name & \textbf{0.0484} & \textbf{0.0605} & 0.0288 & \textbf{0.0327}  \\
\midrule
\multirow{2}{*}{LLaMA-1B} 
& TIGER & 0.0202 & 0.0282 & 0.0202 & 0.0226 \\
& \name & \textbf{0.0403} & \textbf{0.0444} & \textbf{0.0330} & \textbf{0.0342} \\
\midrule
\multirow{2}{*}{LLaMA-3B} 
& TIGER & 0.0242 & 0.0282 & 0.0192 & 0.0205 \\
& \name & \textbf{0.0403} & \textbf{0.0403} & \textbf{0.0373} & \textbf{0.0373} \\
\bottomrule
\end{tabular}}
\end{table}

\subsection{Backbone Adaptability}
We further evaluate the adaptability of our semantic tokenizer by applying the item IDs generated by \name\ to finetune different LLM backbones for sequential recommendation, as shown in Table~\ref{tab:backbone}. Across all backbones, \name\ consistently outperforms TIGER, demonstrating that its structured tokenization provides a more informative and transferable representation for downstream models.
While TIGER achieves good performance under the T5 backbone, its effectiveness drops noticeably when scaling up to larger decoder-only architectures such as LLaMA-1B and LLaMA-3B.  
In contrast, \name\ maintains stable and competitive results across all backbones, demonstrating that its coarse-to-fine tokenization provides a more consistent and transferable representation.  



\section{Related Work}
In this section, we review related work in two key areas: generative recommendation and tokenization for recommendation.

\subsection{Generative Recommendation}

Traditional sequential recommender systems often rely on large item embedding tables~\citep{he2020lightgcn, kang2018self, mnih2007probabilistic}, which pose scalability and memory challenges. In contrast, generative recommendation~\citep{geng2022recommendation, rajput2023recommender, metaHSTU, yue2023llamarec, hou2025actionpiece} encodes items as tokens from a shared vocabulary and generates the next item in an autoregressive manner, enabling open-ended, scalable, and compositional recommendation. Recent methods have explored this paradigm from various angles. P5~\citep{geng2022recommendation} formulates recommendation as a language generation task, prompting pretrained language models using instruction templates and textual item identifiers. TIGER~\citep{rajput2023recommender} represents each item as a semantic ID composed of discrete codes via residual vector quantization and generates them using a Transformer decoder. IDGenRec~\citep{tan2024idgenrec} generates short, interpretable token sequences directly from item metadata. These works demonstrate the potential of generative recommenders in terms of scalability~\citep{rajput2023recommender, lin2024data, xu2024slmrec}, generalization~\citep{rajput2023recommender,huang2024large}, and LLM alignment~\citep{jin2023language}. However, most methods represent each item as a flat ID or sequence of tokens without internal structure, making the generation process less controllable. This also increases the risk of pruning correct items early during beam search, due to the lack of intermediate semantic cues. In contrast, our method introduces a hierarchical tokenization scheme that preserves multi-level item semantics and aligns naturally with users preference refinement.

\vspace{-1em}
\subsection{Tokenization for Recommendation}

A core challenge in LLM-based recommendation is how to tokenize items and interactions in a way that balances semantic fidelity, behavioral alignment, and modeling flexibility. Prior work has explored various tokenization paradigms. ID-based methods~\citep{geng2022recommendation, hua2023index, chu2023leveraging} offer simplicity and efficiency, but lack semantic expressiveness. Textual tokenization~\citep{zhang2023recommendation, zheng2024adapting, lin2024bridging, tan2024idgenrec, bao2025bi} leverages pretrained language models but often struggles with structural consistency and noisy representations. Vector quantization-based approaches~\citep{rajput2023recommender, zheng2024adapting, wang2024learnable, lin2025order, hou2025actionpiece,deng2025onerec,han2025mtgr} discretize continuous embeddings into compact token sequences, enabling compatibility with autoregressive generation. Despite their merits, most of these approaches collapse item semantics into a flat representation, discarding hierarchical structure and making it difficult to integrate collaborative signals meaningfully. Some recent methods attempt to inject CF information~\citep{wang2024learnable, zhang2023collm}, yet they generally overlook positional structure and coarse-to-fine conditioning. P5-SemID~\citep{hua2023index} proposes hierarchical identifiers using taxonomies, but they are non-learnable and lack discriminative power or CF awareness. In contrast, our approach introduces a learnable, structured tokenization framework that constructs a coarse-to-fine semantic hierarchy from item metadata, applies coarse-to-fine quantization, and integrates CF signals as an additional fine-grained token. Our method is conceptually related to VAR~\citep{tian2024visual}, which performs hierarchical generation in vision, but differs in its application to discrete, multi-level textual metadata. Unlike the continuous latent vision maps in VAR, our tokenization handles discrete semantic structures, requiring explicit construction and level-wise quantization and generation tailored for recommendation tasks.

\vspace{-1mm}
\section{Conclusion}
In this work, we proposed \name, a generative recommendation framework that models item semantics in a coarse-to-fine manner through structured tokenization and hierarchical generation. By preserving the multi-level structure of item metadata and integrating collaborative signals, \name\ enables more accurate and interpretable recommendations. Both theoretical analysis and empirical results confirm its effectiveness across diverse datasets and model backbones, highlighting the importance of structured token design in generative recommendation. Beyond improving performance, \name\ provides a new perspective on how generative models can capture user intent evolution through hierarchical semantics.

\bibliographystyle{assets/plainnat}
\bibliography{ref}

\clearpage
\appendix
\noindent \textbf{\Large Appendix}



\appendix

\section{Proof for Proposition 1}
\label{appendix:proof}

\textbf{Proposition 1.}
\label{proposition}
Let \(\mathbb{E}_\text{hier}[d(i,\hat\imath)]\) and \( \mathbb{E}_\text{indep}[d(i,\hat\imath)] \) denote the expected dissimilarity between the predicted item \( \hat{\imath} \) and the ground truth item \( i \), under hierarchical and independent decoding respectively. 
Then,
\begin{equation}
    \mathbb{E}_\text{hier}[d(i,\hat\imath)] < \mathbb{E}_\text{indep}[d(i,\hat\imath)].
\end{equation}

\textit{Proof.} 
We model hierarchical decoding as a traversal over a complete \( V \)-ary tree of depth \( K \), where each internal node represents a prefix (partial code), and each edge corresponds to selecting a token from a vocabulary of size \( V \). Each item is uniquely represented by a path from the root to a leaf, consisting of \( V \) sequential token choices. At each level, the probability of correctly predicting the next token (conditioned on all previous ones) is $p$.

\textbf{Hierarchical decoding.}
The decoder predicts each token level by level. If it fails at level $k$, the dissimilarity is defined as $K-k$. The probability of success at level $k$ is $p^k$, and complete success has probability $p^K$. Hence, the expected dissimilarity is:

\begin{equation}
\mathbb{E}_\text{hier}[d(i,\hat\imath)] = K - \sum_{k=1}^K p^k = K - p - p^2 - \cdots - p^K.
\end{equation}

\textbf{Independent decoding.}
Each of the $V$ tokens is predicted independently, without leveraging the tree structure. This corresponds to randomly sampling a code from the $V^K$ possible sequences, where each leaf (i.e., each item) has equal probability. 
Hence,
\begin{equation}
\mathbb{E}_\text{indep}[d(i,\hat\imath)] = \left(K - \sum_{k=1}^{K} \frac{1}{V^k} \right) \cdot \frac{ 1 - p^K }{1 - \frac{1}{V^K} }.
\end{equation}

Define an auxiliary function:
\begin{equation}
\psi(p) := \frac{K - \sum_{k=1}^K p^k}{1 - p^K},
\end{equation}
which we will use to simplify the expected dissimilarity. In the following \textbf{Lemma 1}, we will show that the function \( \psi(p) \) is strictly decreasing with respect to \( p \) over the interval \( [0,1] \).

Since $p>1/V$, then the ratio between hierarchical and independent decoding dissimilarity is smaller than 1:
\begin{align}
\frac{\mathbb{E}_\text{hier}[d(i,\hat\imath)]}{\mathbb{E}_\text{indep}[d(i,\hat\imath)]} &= \frac{K - \sum_{k=1}^K p^k }{\left(K - \sum_{k=1}^{K} \frac{1}{V^k} \right) \cdot \frac{ 1 - p^K }{1 - \frac{1}{V^K} }}
\\&=\frac{\frac{K - \sum_{k=1}^K p^k}{1 - p^K}}{\frac{K - \sum_{k=1}^{K} \frac{1}{V^k}}{1 - \frac{1}{V^K}}}=\frac{\psi(p)}{\psi\left(\frac{1}{V}\right)}<1.
\end{align}
It follows that $\mathbb{E}_\text{hier}[d(i,\hat\imath)] < \mathbb{E}_\text{indep}[d(i,\hat\imath)]$.

\paragraph{Conclusion.} 
This result highlights the benefit of hierarchical decoding: correct predictions at higher levels constrain the prediction space at lower levels, resulting in smaller expected dissimilarity between predictions and the ground truth. Independent decoding lacks this structural bias and thus incurs a higher error on average.

\begin{lemma}
The function $\psi(p):=\frac{K - \sum_{k=1}^K p^k}{1 - p^K}$ is strictly decreasing over $[0, 1]$ whenever $K>1$. 
\end{lemma}

\textit{Proof.} We rewrite the function as:
\begin{equation}
\psi(p) = \frac{K - \sum_{k=1}^K p^k}{1 - p^K} = 1+\sum_{k=1}^{K-1}\frac{1-p^k}{1-p^K}.
\end{equation}
It suffices to show that for every $1\le k\le K-1$,
\begin{equation}
\psi_k(x):=\frac{1-p^k}{1-p^K}
\end{equation}
is strictly decreasing. To analyze its monotonicity, we check the sign of its derivative:
\begin{equation}
\psi_k'(x)=-\frac{p^{k-1}}{(1-p^K)^2}(k+(K-k)p^K-Kx^{K-k}).
\end{equation}
Since $-\frac{p^{k-1}}{(1-p^K)^2}<0$ over $(0,1)$, it remains to analyze the sign of
\begin{equation}
\zeta_k(p):=k+(K-k)p^K-Kp^{K-k}.
\end{equation}
Since for any $0<x<1$,
\begin{equation}
\zeta_k'(p)=-(K-k)Kp^{K-k-1}(1-p^k)<0,
\end{equation}
then $\zeta_k(p)$ is strictly decreasing over $(0,1)$. Thus, for any $0<p<1$,
\begin{equation}
\zeta_k(p)>\zeta_k(1)=k+(K-k)-K=0.
\end{equation}
Hence, for any $0<p<1$,
\begin{equation}
\zeta_k'(x)=-\frac{p^{k-1}}{(1-p^K)^2}\zeta_k(p)<0.
\end{equation}
Therefore, for every $1\le k\le K-1$, $\psi_k(p)$ is strictly decreasing over $p\in[0,1]$. It follows that
\begin{equation}
\psi(p)=1+\sum_{k=1}^{K-1}\psi_k(p)
\end{equation}
is strictly decreasing over $[0,1]$.

\section{Analysis on Code Collision}
\label{app:code_collison}
We analyze the impact of hierarchical tokenization on code collisions. Table~\ref{tab:collision} compares CoFiRec with the RQ-VAE tokenizer under the same codebook size (256 per level). CoFiRec achieves significantly lower collision rates across all datasets, reducing collisions by more than one order of magnitude. This demonstrates that decomposing item semantics into multiple hierarchical levels and incorporate collaborate signals as the most fine-grained token granularity allows more efficient utilization of the code space and avoids over-compression within a single latent space.

To further investigate how the size of the codebook affects collisions, we evaluate three configurations that correspond to the settings in Table~\ref{tab:codebook}. As shown in Table~\ref{tab:collision-size}, enlarging the codebook capacity in deeper semantic levels further suppresses collisions. The collision rate remains near zero under the medium and large configurations, confirming that granularity-aware capacity allocation effectively mitigates redundancy in the token space.

\begin{table}[h!]
\centering
\small
\caption{Code collision rate comparison between CoFiRec and RQ-VAE tokenizers.}
\label{tab:collision}
\begin{tabular}{lccc}
\toprule
\textbf{Model} & \textbf{Instruments} & \textbf{Yelp} & \textbf{Beauty} \\
\midrule
RQ-VAE Tokenizer & 0.0027 & 0.0410 & 0.0009 \\
CoFiRec Tokenizer & \textbf{0.0001} & \textbf{0.0002} & \textbf{0.0002} \\
\bottomrule
\end{tabular}
\end{table}

\begin{table}[h!]
\centering
\small
\caption{Code collision rate of CoFiRec under different codebook size configurations on the Instruments dataset.}
\label{tab:collision-size}
\begin{tabular}{lcc}
\toprule
\textbf{Codebook Size} & \textbf{Level Size} & \textbf{Collision Rate} \\
\midrule
Small  & [256, 256, 256, 256] & 0.0001 \\
Medium & [256, 256, 512, 512] & 0.0000 \\
Large  & [256, 256, 512, 1024] & 0.0000 \\
\bottomrule
\end{tabular}
\end{table}

\section{Automatic Hierarchy Construction (AHC)}
\label{app:AHC}
In our main experiments, \name\ is built upon simple and widely available fields—\textit{category}, \textit{title}, and \textit{description}—which exist in most practical product recommendation scenarios. To verify the generality of \name, we further explore whether meaningful hierarchical semantics can be constructed when structured metadata is unavailable. Specifically, we use GPT-4o to automatically generate 4-level hierarchies for items in the Instruments dataset based solely on their textual information. The generated hierarchies are used only to test the feasibility of our framework under minimal metadata conditions, rather than to replace real-world metadata.  
The GPT-based variant, denoted as AHC (Automatic Hierarchy Construction), serves as a fallback strategy for extreme cases where structured metadata is missing. Shown in Table~\ref{tab:AHC}, AHC still achieves competitive performance.  

\begin{table}[h!]
\centering
\caption{ Performance comparison on the Instruments dataset between TIGER, \name\ with full metadata, and \name\ with automatically constructed hierarchies (AHC). }
\label{tab:AHC}
\resizebox{0.47\textwidth}{!}{%
\begin{tabular}{lcccc}
\toprule
\textbf{Method} & \textbf{R@5} & \textbf{R@10} & \textbf{N@5} & \textbf{N@10} \\
\midrule
TIGER & 0.0444 & 0.0484 & 0.0293 & 0.0307 \\
\name \ w/ AHC  & 0.0444 & 0.0484 & \textbf{0.0313} & 0.0327 \\
\name & \textbf{0.0484} & \textbf{0.0605} & 0.0288 & \textbf{0.0327} \\
\bottomrule
\end{tabular}%
}
\end{table}

\section{Future Work}
Building upon the promising results of CoFiRec, several future directions can further extend its capabilities and impact:

(1) \textbf{End-to-end integration of tokenization and generation.}  
Currently, the semantic tokenizer and the generative recommendation model are trained separately. While this modularity facilitates transferability across tasks, a fully end-to-end framework could enable stronger mutual adaptation between tokenization and generation. Future work may investigate joint optimization strategies to align the latent token space with downstream generation objectives, improving both efficiency and semantic consistency.

(2) \textbf{Extension to multimodal recommendation.}  
Our present framework operates on textual metadata, but many real-world platforms involve rich multimodal signals such as product images, videos, or user interactions. Extending CoFiRec’s coarse-to-fine tokenization to incorporate visual and other modalities offers a natural next step. Such an expansion could further enhance semantic grounding and lead to more comprehensive generative recommendation systems.

\section{Prompt for Automatic Hierachy Construction}
\label{app:prompt}

\begin{tcolorbox}[
  title=Example: Generating Semantic Hierarchy with GPT,
  coltitle=white,
  colbacktitle=gray,
  colback=gray!3,
  colframe=black,
  fonttitle=\bfseries,
  arc=2mm,
  boxrule=0.5pt,
  enhanced jigsaw,
  width=\linewidth,
  left=6pt,
  right=6pt,
  top=6pt,
  bottom=6pt
]

\textbf{Input Prompt:}

You are a product taxonomy expert specializing in musical instruments.

Given an item’s title and description, generate a 4-level coarse-to-fine hierarchical categorization within the "Instruments" domain. The hierarchy should reflect increasing specificity from general product type (Level 1) to full product-level detail (Level 4). Do not include ``Musical Instruments'' or ``Instruments'' as Level 1 -- assume the item is already within that category.

\textbf{Guidelines:}
\begin{itemize}
    \item Each level must be a single short phrase or sentence.
    \item Each finer level should include or imply the coarser level’s meaning.
    \item Level 4 should include brand, model, main features, and supported use cases, written as a concise product summary.
    \item Format the output as: Level 1: \dots, Level 2: \dots, Level 3: \dots, Level 4: \dots
\end{itemize}

\textbf{Here is the item:}
\begin{itemize}
    \item \textbf{Title}: "Chrome Oval Style 1/4 Guitar Pickup Output Input Jack Plug Socket for Fender Strat."  
    \item \textbf{Description}: "Specification: 100\% Brand New, Never Used This is a mono Electric guitar output jack plate.  Width at the widest point of the plate: 2.7cm Has 3 Total Prongs Measurement between the centre of the 2 screw holes: 3.8cm Replace your worn out parts Boat Style (Oval Indented) Standard recessed oval plate which is used commonly on some LP SG guitars Color:chrome  Package Include: 1x Chrome Jack Plate."
\end{itemize}

\vspace{1em}

\textbf{Hierarchy Output:}
\begin{itemize}
  \item \textbf{Level 1:} Guitar Accessories
  \item \textbf{Level 2:} Guitar Hardware Components
  \item \textbf{Level 3:} Guitar Output Jack Plates
  \item \textbf{Level 4:} Chrome Oval Style 1/4 Guitar Pickup Output Input Jack Plug Socket for Fender Strat, featuring a mono electric guitar output jack plate with a standard recessed oval design, suitable for replacing worn-out parts on LP and SG guitars.
\end{itemize}

\end{tcolorbox}

\section{Dataset Statistics}
\label{app:dataset}
\begin{table}[b!]
\centering
\small
\caption{Dataset statistics for Instruments, Yelp, and Beauty.}
\label{tab:dataset-stat}
\resizebox{0.7\columnwidth}{!}{
\begin{tabular}{lcccc}
\toprule
\textbf{Dataset} & \textbf{\# Users} & \textbf{\# Items} & \textbf{\# Interactions} & \textbf{Avg. Seq. Length} \\
\midrule
Instruments & 24{,}772 & 9{,}922 & 206{,}153 & 8.32 \\
Yelp        & 30{,}431 & 20{,}033 & 316{,}354 & 10.40 \\
Beauty      & 22{,}363 & 12{,}101 & 198{,}502 & 8.88 \\
\bottomrule
\end{tabular}}
\end{table}

We summarize the datasets used in our experiments in Table~\ref{tab:dataset-stat}.  
The three datasets—\textbf{Instruments}, \textbf{Yelp}, and \textbf{Beauty}—are derived from the Amazon review and Yelp dataset~\citep{yelpdataset} and are widely adopted in sequential recommendation research.  
Each dataset consists of user–item interaction sequences, where for each user $u$, the sequence is denoted as $\mathcal{S}_u = [i_1, i_2, \dots, i_T]$, and the goal is to predict the next item $i_{T+1}$.  
All datasets are preprocessed by filtering out users and items with fewer than five interactions and sorting interactions chronologically.

\section{Runtime and Scalability Analysis}
\label{app:runtime}

We analyze the computational efficiency and scalability of \name\ compared with TIGER.  
Since access to large-scale industrial datasets is limited, we follow representative generative recommendation studies (e.g., TIGER, P5, IDGenRec) and conduct experiments on standard public datasets such as Amazon and Yelp.  

To evaluate scalability, we measure both training and inference runtimes under the same experimental environment using the same backbone.  
As summarized in Table~\ref{tab:runtime}, \name\ introduces minimal overhead compared to TIGER across both the tokenization and generation stages.  
The additional hierarchical modules incur only slight increases in computation time during tokenization, while generation remains nearly identical in efficiency.

\begin{table}[t!]
\centering
\small
\caption{Runtime comparison between \name\ and TIGER.}
\label{tab:runtime}
\resizebox{0.6\columnwidth}{!}{
\begin{tabular}{llcc}
\toprule
\textbf{Method} & \textbf{Stage} & \textbf{Training Time} & \textbf{Testing Time} \\
\midrule
TIGER   & Tokenization & 0.3488s / step & 0.3113s / step \\
\name   & Tokenization & 0.3584s / step & 0.3256s / step \\
TIGER   & Generation   & 400.6869s / epoch & 0.0998s / step \\
\name   & Generation   & 402.3185s / epoch & 0.0943s / step \\
\bottomrule
\end{tabular}}
\end{table}

In terms of memory usage, \name\ remains highly efficient: the tokenization stage requires less than 1.8 GB of GPU memory, and generation under 4 GB when fine-tuning the same T5 backbone—comparable to TIGER.  
The main computational cost still lies in downstream fine-tuning of large language models, which is a shared characteristic across all generative recommender systems.  

\end{document}